\documentclass{article}%
\usepackage{amsfonts}
\usepackage{amsmath}
\usepackage{amssymb}
\usepackage{graphicx}%
\begin{document}

\title{The power of AQFT: the area law for entropy of localized quantum matter\\{\small submitted contribution to Fifth International Conference on
Mathematical Methods in Physics April 2006, Rio de Janeiro}}
\author{Bert Schroer\\CBPF, Rua Dr. Xavier Sigaud 150 \\22290-180 Rio de Janeiro, Brazil\\and Institut fuer Theoretische Physik der FU Berlin, Germany}
\date{October 2006}
\maketitle

\begin{abstract}
The algebraic approach to QFT, which for several decades has enriched QFT with
structural theorems, has recently shown its utility in various constructions
of actual interest. In these lecture notes I explain how AQFT (in particular
the modular theory of operator algebras) implies paradigmatic conceptual and
mathematical changes while fully preserving the physical principles which
underly QFT. As an illustration of actual interest I use holography on
null-surfaces and the ensuing area law for entropy of localized matter in the
vacuum state.

\end{abstract}

\section{A post-standard model quantum field theory?}

Looking at the present particle physics scene one may easily get the somewhat
misleading impression that besides string theory and loop quantum gravity (and
perhaps leaning back and just waiting for new results from the LHC machine in
construction)\ there are no other worthwhile fundamental alternatives

Here I would like to point out the existence of \textit{a third way} whose
pursuit neither requires to wait for the moment when new experimental data
resolve the Higgs issue or supersymmetry (which the impatiently expected LHC
data may not even be able to deliver), nor does it involve the risks of a
community getting lost in the blue yonder of the aforementioned highly
speculative ideas.

This third way was always available but in the good times of the past, when
ideas supported by existing computational methods led to a lot of progress,
there was no strong motivation to think about it. It is based on the
observation that QFT in contrast to QM has remained a conceptually unfinished
project. Our knowledge has mainly been obtained by canonical or functional
integral quantization of classical objects, i.e. by a parallelism to a world
of classical field theory\footnote{Often invented for the only purpose of
being able to grind it through the quantization mill.} or in the words of the
protagonist of field quantization Pascual Jordan, with the help of "classical
crutches" \cite{Khar}.

We know about the existence of an autonomous world of QFT beyond classical
analogs \cite{Haag} from the result of some courageous explorers who, without
much encouragement from people in the particle theory mainstream who were busy
looking for new discoveries with the standard methods, have provided us with a
wealth of structural insights into QFT. From very recent constructions of
certain low-dimensional QFT models, which do not permit the attachment of a
Lagrangian name \cite{Le}\cite{Ba-Fo-Ka}, we also know that there are many
more models than Lagrangian interactions; in fact the existence of a
Lagrangian name is neither sufficient for insuring its existence nor necessary
in order to know its physical content.\ With other words Lagrangian names do
not make models more amenable for mathematical control, conceptual
understanding of its physical content or nonperturbative computational access.
The only thing which we can claim with considerable confidence is that the SM
is most probably the best Lagrangian straight jacket (enriched with group
symmetries and the gauge formalism) for the subatomic world which we could
have hoped for.

On order to make progress beyond the SM one should therefore \textit{explore
QFT beyond its narrow Lagrangian quantization setting} using the very firm
guidance of its underlying principles. In practical terms it means bringing
the computational side of scientific production in particle theory into a
better balance with conceptual developments. The number of open
well-formulated theoretical probIems is presently larger than at any other
time in the history of particle theory (but one only becomes aware of this
once one steps outside the lure of present fads). As a result of decades of
uninterrupted success of particle physics in the last century there was little
incentive to take up the unsolved conceptual problems which were left on the
wayside; it is precisely this past success which makes a conceptual
re-orientation at this late time difficult and time consuming. There is the
widespread feeling that the relative ease with which the standard model was
discovered is somehow related to the present difficulties to go beyond it; in
more concrete terms one suspects that the conceptual mathematical setting
which was essential for its formulation (i.e. Lagrangian quantization enriched
with group representation theory) does not contain the right instruments
needed for answering the remaining questions. This would amount to an
admission that the success of renormalized perturbation theory and its finest
product, the standard model, only temporarily sidelined the plea for a total
autonomous framework of QFT and that the right time to return to this old
issue with new vigor is now (even before LHC is running). The long lasting
crisis in particle physics and the many failed attempts in trying to cure the
symptoms with more of the same medicine forces us to take a new look at this
old problem. In \cite{Crisis} I argued that self-interacting massive
vector-mesons are only compatible with locality if they are accompanied by
additional physical degrees of freedom whose simplest realization is a scalar field.

Looking into the QFT books one may have the impression that after Wigner's
successful attempt to present the classification of particle spaces without
reference to quantization (and its use in e.g. Weinberg's text book only for
additional support of the Lagrangian setting without investigating its
potential for exploring Jordan's dream of an \textit{autonomous QFT without
classical crutches}) not much has happened. Fortunately this first impression
is not correct; there has been indeed steady progress in the pursuit of an
autonomous presentation of QFT. Since in most textbooks the terminology "QFT"
has been identified with the Lagrangian approach, it became customary practice
to use the terminology LQP (local quantum physics, thus indicating that causal
localization is its central concept) or AQFT (algebraic QFT, using operator
algebra theory), when authors want to highlight that they are using a wider
setting of QFT while maintaining its physical principles.

In a recent critical essay (last section in \cite{Crisis}) on the present
situation I have argued that the new way of looking at particle physics sheds
a new and different light on such important subjects as gauge theory and the
Schwinger-Higgs screening mechanism\footnote{This is often incorrectly called
the Higgs \ "symmetry-breaking".} for massive vectormesons. Some of these new
perspectives will be presented in Mund's contribution to this conference and
published in the same proceedings\footnote{The title of J. Mund's contribution
for the proceedings will be \textit{String-Localized Quantum Fields and
Modlular Localization.}}. There is also the rapidly changing (under the
influence of AQFT) setting of QFT in CST and an emerging new perspective on
perturbative gravity.

Here I will use the very interesting ideas about black hole thermal aspects as
a motivating vehicle to present results about holography onto null-surfaces
\cite{AreaI}\cite{AreaII} which draws heavily on operator algebra methods (in
particular for the derivation of thermal aspects of localization and the area
law for the entropy of localized quantum matter). The following small section
about modular theory provides some remarks about the mathematical operator
algebra formalism.

\section{Modular theory and QFT in terms of positioning of monads}

In the algebraic approach the starting point for structural investigations of
Poincar\'{e} invariant QFT has been a spacetime-indexed net of operator
algebras which is required to fulfill certain physically motivated and
mathematically well-formulated properties related to causal locality and
stability of relativistic quantum matter (energy positivity, KMS for thermal
states). It is fairly easy to show that the individual local operator algebras
(in contradistinction to the generated global algebra in a vacuum
representation) do not contain operators which annihilate the vacuum
(separability) but nevertheless generate a dense set of states from the vacuum
(cyclicity) which changes together with their localization region.

This so-called Reeh-Schlieder property turns out to be the prerequisite for
applying the Tomita-Takesaki theory \cite{Haag}; the latter is an extremely
rich and profound mathematical theory within the setting of operator algebras.
It associates with a pair of a weakly closed operator algebra (von Neumann
algebra) acting cyclic and separating on a reference state vector
($\mathcal{A},\Omega$) to two \textit{modular}\footnote{The theory is a vast
extension of the notion of (uni)modularity of Haar measures in group algebras
to the general setting of von Neumann algebras and their classification
\cite{Su}.} \textit{objects}: a one-parametric automorphism group of
$\mathcal{A}$ and a TCP-like anti-unitary reflection $J$ (called the modular
involution) whose adjoint action maps $\mathcal{A}$ into its commutant
operator algebra $\mathcal{A}^{\prime}$. This theory unfolds its conceptual
power in a situation of several such algebras within a common Hilbert space.
In fact placing a finite number of such operator algebras into certain
relative positions\footnote{The requirement is that the positioning should be
natural within the logic of modular operator theory (using the concept of
modular inclusions and modular intersections). For higher than 3 spacetime
dimensions the presently known descriptions still look somewhat concocted
\cite{Ka-Wi}.}, the various individual modular automorphism groups generate
noncompact spacetime symmetry groups and the action of the latter on the
original finite number of operator algebras generates a \textit{net of
operator algebras} with the properties demanded by AQFT \cite{Ka-Wi}. The
opposite way around, each AQFT permits a representation in terms of a modular
positioning of a finite number of such operator algebras. In case of a
Moebius-invariant chiral QFT the number is two, for 3-dimensional QFT one
needs 3 algebras and 4-dimensional QFT can be generated from 6 appropriately
positioned algebras: the number increases with increasing spacetime dimension.
A closer look reveals that the so positioned algebras are necessarily
\textit{isomorphic copies} of one and the same operator algebra which
according to Connes refinement of the Murray-von Neumann classification is the
\textit{unique hyperfinite type III}$_{1}$\textit{\ factor algebra \cite{Yng}
}\ often called \textit{monade} because of its unique constructive role in QFT
which follows precisely the idea of Leibnitz about the basic properties which
constitute physical reality.

This is not the place to give a mathematical account of the role of modular
theory within the operator-algebraic formulation of QFT. But for conveying
some aspects of its revolutionary new perspective a few remarks about its
conceptual-philosophical content are in order.

If one identifies this distinguished unique hyperfinite type III$_{1}$ von
Neumann factor algebra as indicated with the role of \textit{monades} in
Leibniz's construct of ideas about what constitutes reality, one finds a
perfect match: \textit{the physical reality of relativistic local quantum
matter in Minkowski spacetime originates from the (modular) positioning of a
finite number of copies of one abstract monade}; no sense of individuality can
be attributed to the monade, apart from the fact that those hyperfinite
factors allow inclusions and intersections (which are meaningless for points)
it is as void of individual structure as a point in geometry (if you have seen
one monade, you know them all). The rich content of a quantum field
theoretical model including its physical interpretation (particles, spacetime
and inner symmetries, scattering theory....) is solely encoded in the relative
positions of a finite number of its monades \cite{Ka-Wi}. In view of the
conceptual simplicity and the radical paradigmatic aspect of this totally
autonomous and rigorous setting for characterizing particle physics, there is
no way in which one could think of QFT as a closed theory. There is simply too
much unexplored terrain beyond the Lagrangian quantization approach (which in
view of its parallelism to classical physics cannot really be considered as an
autonomous setting). The constructive use of this novel characterization of
QFT is still terra incognita but the fact that it yields an alternative
description of QFT is beyond any doubt.

The reader who is familiar with Vaughn Jones's subfactor theory \cite{sub}
will notice that the underlying philosophy of subfactor theory illustrates
this monade picture, the only difference is that as a result of a different
aim the subfactor theory is in many aspects simpler. During the last 3 decades
subfactor theory has become a mathematical very mature and rich theory. The
underlying motivation in this case was to generalize the theory of finite and
compact groups by passing from Galois's inclusion of commutative polynomial
fields to noncommuting operator algebras. To achieve this it suffices to
identify the monade with the simpler hyperfinite type II$_{1}$ factor on which
a tracial state can be defined. This monade is too small for obtaining
spacetime symmetries and localization- as well as thermal- properties; the
implementation of these properties necessitate the use of previous field
theoretic monades (as well as the replacement of \textit{Jones inclusions} by
\textit{modular inclusions}). The theory of Jones inclusions is older and much
more developed than the notion of modular inclusion (which is the important
concept in the above \textit{monade presentation of QFT}) and has served as a
source of inspiration to AQFT. It was preceded by the DHR theory of localized
endomorphism which in the Doplicher-Haag Roberts work was the crucial
operator-algebraic concept to unravel the spacetime localization origin of
statistics and global inner symmetries in QFT \cite{Haag}.

It is interesting to look at the way in which this modular theory based
setting of QFT leads to spacetime symmetry, localization and causality in
Minkowski spacetime i.e. to ask the question what is the algebraic germ for
the rich geometrical aspects of QFT. It turns out that this is related to the
previously mentioned positions of the dense subspaces which are cyclically
generated by acting with the monade on the vacuum. The relation is made
precise by modular theory in terms of the domain of the unbounded involutive
Tomita operator S. This encoding of geometric aspects into abstract domain
properties of unbounded operators is characteristic of Tomita operators; no
other operator is capable to e.g. lead from the inclusion of domains of
unbounded operators to inclusions of geometric localization regions in
Minkowski spacetime. This way of getting from physical principles of local
quantum physics to geometric properties is very different from say the better
known Atiyah-Witten-Segal setting which uses the suggestive power of the
classical geometric and topological aspects of euclidean functional integral
representations for mathematical innovations. The mathematical results
obtained in this way are independent of whether the motivating physical
setting is metaphoric or autonomous\footnote{For the derivation of the e.g.
Jones polynomial it is of no importance whether the Euclidean Chern-Simons
action fits into the Osterwalder-Schrader framework i.e. is associated with a
physical correlation function (it is not!) as long as its mathematical meaning
is well-defined and its quasiclassical approximation is under mathematical
control. In such metaphoric QFT-math connections it is also of no interest
whether objects "derived" in this way really fulfill the starting relation or
not.} (Euclidean functional integrals are benign metaphors i.e. the answers
obtained with physical hindsight are correct even though they violate the
functional integral representations from which they allegedly were derived).

AQFT is very different in its aims, it starts from mathematically rigorously
formulated physical principle and derives consequences by operator algebraic
methods. Using this approach in low dimensional QFT one arrives e.g. at braid
group statistics \cite{F-R-S} in low-dimensional QFT in a way which is
reminiscent of Jones subfactor theory. Whereas the quasiclassical
approximation on functional integrals for Chern-Simons actions leads to Jones
polynomials without revealing their physical interpretation in terms of
particle statistics, the physical origin and interpretation in the algebraic
approach remains clear throughout all computational steps.

One should perhaps mention that the discovery of modular operator theory was
made independently at the same time by mathematicians (Tomita, with
considerable enrichments by Takesaki) as well as by physicists
(Haag-Hugenholtz-Winnink). Naturally physicists do not aim at the greatest
mathematical generality but rather introduce mathematical concepts which are
designed to solve a specific physical problem. The physical problem which
brought H-H-W into the conceptual proximity of modular theory was the
formulation of quantum statistical mechanics for open systems (i.e. directly
in the thermodynamic limit) \cite{Haag}. I do not know any other case in the
history of mathematical physics where a fundamental mathematical theory was
simultaneously (and independently) discovered by mathematicians and
physicists. After both sides realized this the concepts and terminology
quickly merged together. Even in the historical example of quantum mechanics,
the Hilbert space theory was already fully available and it took the
physicists some years to become aware of it. The relation of AQFT to the
mathematical operator algebra theory is very different from the more
fashionable but inherently metaphoric mathematics-particle physics relation.

\section{Modular theory, classification and construction of models}

The fact that the above presentation of Poincar\'{e}-invariant QFT in terms of
positioning of monades is under rigorous mathematical control does not yet
mean that it can be readily used for classifying and constructing models. To
facilitate such a use it is advantageous to reformulate the monade assumption.
One can show that the assumption of having a unitary representation of the
Poincar\'{e} group and knowing its action on just one monade\footnote{This
means that one knows the position of the wedge algebra within the full algebra
B(H) of all operators i.e. the inclusion $\mathcal{A}(W)\subset B(H).$} is
equivalent to the previous relative placement of several copies. Some
additional thinking reveals that identifying this monade algebra with a
wedge-localized operator algebra in the QFT to be constructed is a good
starting point, because the modular objects of a wedge algebra have a
well-known physical interpretation. Hence knowing a wedge algebra (i.e. its
position in B(H)), the action of the Poincar\'{e} group on it immediately
leads to the knowledge of all wedge algebras; taking suitable intersection of
these wedge algebras and unions of such intersections one obtains the full net
of all spacetime indexed algebras i.e. the QFT associated with the original
wedge-localized monade together with its covariant transformation property. If
the intersections are the trivial algebra (i.e. complex multiples of the
identity) then the wedge algebra has no associated QFT. From a practical point
of view one does not compute directly with wedge algebras but rather with a
generating system of wedge-localized operators which carry a known
representation of the Poincar\'{e} group.

This rather abstract construction idea can be made to work under the quite
strong restriction that there exist wedge-localized generators which, similar
to free fields, upon their one-time application the vacuum create one particle
states without the admixture of particle-antiparticle vacuum polarization
clouds \cite{PFG}. In case such vacuum-\textbf{p}olarization-\textbf{f}%
ree-\textbf{g}enerators (PFG) exist for subwedge spacetime regions (i.e. for
regions whose causal completion is smaller than a wedge) it can be shown (by a
slight generalization of the Jost-Schroer theorem \cite{Wi-St}) that the
theory is free in the sense of being generated by free fields. In some sense
which can be made precise, wedge-localized PFG's are the best localized
objects in interacting theories for which the field theoretic localization
property and that of Wigner particles still coexist simultaneously; for
subwedge localization the inexorable presence of interaction-caused vacuum
polarization prevents the creation of only one particle states (without an
particle/antiparticle polarization cloud) from the vacuum.

A detailed study of PFG wedge-localized generators with translation-invariant
domains \cite{B-B-S} reveals that they only exist in theories with a purely
elastic S-matrix which is only possible in d=1+1 dimensions. It turns out that
the resulting field theories are precisely those of the bootstrap-formfactor
program \cite{Ka} and the wedge-localized PFG's are the Fourier transforms of
the Zamolodchikov-Faddeev algebra generators (which in this algebraic setting
receive a spacetime interpretation). Without the use of modular theory it was
not possible to show that the calculated formfactors really belong to a
well-defined QFT. This existence problem of QFT in this context of factorizing
models has meanwhile been solved \cite{Bu-Le}\cite{Le} by showing that the
double-cone intersections of wedge-localized intersection are really
nontrivial\footnote{The proof based on modular theory still needs to be
extended to the presence of bound states.}. Although these models have no real
particle creation via scattering, their formfactors (matrix-elements of
localized operators) exhibit a very rich vacuum polarization structure. Since
they are analytic in a small region around zero coupling strength and since
there are general structural arguments that their off-shell spacetime
correlation functions admit \textit{no} convergent power-series expansion for
small couplings (they are at best only asymptotically convergent), these
models suggest that on-shell quantities may have better perturbative
properties than off-shell quantities. It would be very interesting to use
these models in order to study this interesting problem in detail.

Without wedge-localized PFG's, i.e. outside factorizing models there are some
(presently still rather) vague ideas of how the fact that the S-matrix has the
interpretation of a relative modular invariant\footnote{Tis means that its
modular involution $J$ is related to that of a corresponding free theory
$J_{0}$ through J=J$_{0}S_{scat}$ where $S_{scat}$ is the scattering matrix.}
may be used in model constructions. They are based on the suspicion that the
old S-matrix bootstrap of the 60s failed (even in form of a perturbative
construction) because this somewhat surprising and powerful connection of the
scattering operator with wedge-localized algebras was not noticed; the hope is
that its use (i.g. in a perturbative bootstrap-formfactor program) could have
a chance to improve this situation. I expect that by combining some ideas of
the old S-matrix approach with this recent framework of modular wedge
localization one will obtain completely new insights into the nonperturbative
structure of QFT.

It is my intense impression that the replacement of crossing and its
substitution by Veneziano's duality was the \textit{wrong turn at one of the
most important cross roads of particle physics} and that it is not possible to
make progress in particle physics without a new problematization of the
bootstrap-formfactor idea. In particular I expect that the geometric aspects
of gauge theory which were important to arrive at the formulation of the
standard model will not be useful in order to go beyond the already 30 year
lasting stalemate of the SM.

\section{Some concrete results about holography, localization entropy and
black holes}

AQFT and in particular its concept of modular localization is most important
in situations in which the conventional quantization approach based on
Lagrangians and functional integrals breaks down. Besides the previously
mentioned factorizing models this is the case when one studies properties
which have no natural association with individual fields. One such situation
is the phenomenon of \textit{localization-thermality }and in particular
\textit{localization-entropy}.

Using a rigorous algebraic formulation of \textit{holographic projection} of
localized quantum matter onto the causal horizon of the localization region,
it is possible to compute the entropy generated by the vacuum polarization
which occurs near the horizon. For technical details we refer to
\cite{AreaI}\cite{AreaII}; in the following we will a description of the main
ideas. The perception that the surface of a localization region is the source
of a very strong vacuum-polarization cloud was historically one of the first
observations which showed the distinction between QM and QFT in the most
dramatic way\footnote{Heisenberg became aware of (infinite) vacuum flucuations
when he tried to define a partial quantum charge by integrating the charge
density over a finite volume. In the modern test function formulation of such
problems the vacuum-fluctuation is kept finite by introduving a collor of size
$\varepsilon$ around the surface inside which the vacuum-polarization cloud
can spread; the divergence re-appears for $\varepsilon$ $\rightarrow0.$It
disappears in the "thermodynamic limit".}.

The causal horizon of any (causally complete) region is a
\textit{null-surface} and the holographic projection keeps the global algebra,
but radically changes its local net structure of the algebraic net of the bulk
to that of the horizon. Although the change in the spacetime encoding leads to
a different physical interpretation there is not much intrinsic meaning to say
the theory "lives in the bulk region " or "lives on the horizon" because this
is not a property which is intrinsic to the abstract algebraic substrate but
rather depends on the way in which it is spatially organized.

The much advertised AdS-CFT holography\footnote{The idea that there exists a
separate \ "gravitational holography" (different from the present field
theoretic version which of course permits an extension to QFT in CST) is an
metaphoric illusion. The Maldacena conjecture versus the Rehren theorem on
AdS-CFT has been a subject of heated debates.} is a very special kind of
holographic projection onto a \textit{timelike boundary} (meaning it contains
the timelike direction, this is often called \textit{brane}) at infinity;
since it is an \textit{isomorphism} (one-to one- holographic projection) under
which the conformal spacetime covering group $\widetilde{SO(4,2)}$ remains
invariant. In that case the reprocessing of the (global) bulk spacetime net
structure to that on the CFT boundary is rather straightforward and the name
\textit{correspondence} is more appropriate than holographic projection (which
we will henceforth only use in case of null-surfaces). The AdS-CFT
correspondence is one between QFT with the highest possible vacuum symmetries
on geodetical complete globally hyperbolic manifolds of different spacetime
dimensions; there are many models with lesser number of global symmetries
which permit "partial" isomorphisms between algebras associated with
subregions of different spacetimes with the same spacetime dimensions.

Unfortunately\textbf{ }the idea of holography entered particle physics with a
heavy metaphoric burden as a result of its alleged quantum gravity connection
\cite{Ho}. This prejudice has proven quite resistant and it is the root of a
conjecture the famous Maldacena AdF-CFT conjecture \cite{Mal1} with led to
thousands of papers in its bow wave \cite{Crisis}; it seems that no rational
argument \cite{Dies} is able to convince string theorists that the assumption
on which it is based namely that holography in the above sense (of a change of
spacetime encoding of a fixed algebraic quantum substrate) is capable of
producing a QG theory on AdS (a contradiction in terms since QG is by
definition background independent) is totally metaphoric \cite{Mal2} without
any reasonable chance to ever be backed up by an autonomous physical argument.
The conjecture which is backed up by a pathetically incomplete calculation
outside any mathematical control is interpreted on the AdS side as a kind of
dual QCD (strong QCD coupling limit where one expects confinement). The
contradiction to the above rigorous AdS-CFT holography evaporates if one
changes the terminology and interprets the Maldacena conjecture as one about
an unknown weak coupling AdS theory corresponding to a strongly coupled QCD.
If this would be confirmed by more trustworthy approximations it will be a
sensational result (even if presently it has no conceptual basis at all) which
is totally independent on ST or QG. The ST motivation could then be remembered
as a historical footnote and an insight into nonperturbative QCD could then be
celebrated as the result of a collective effort of hundreds of researchers in
thousands of papers (which would be a completely new sociological phenomenon
in particle physics). 

It seems that as a collateral effect of this kind of over-exposure of a
fashionable subject the impressive progress about the validity of local
covariance (leading to background independence) for quantum matter in CST
which arose from a very nontrivial extension of the Haag-Kastler framework
(including some convincing but not yet rigorous arguments that this continues
to hold in a perturbative approach to the Einstein-Hilbert action) remained
unnoticed \cite{Br-Fr-Ve}\cite{Br-Fr}. 

By far the most interesting case is the holographic projection on
null-surfaces (not a correspondence!) because it permits to focus on aspects
which are pretty much out of reach within the spacetime organization which the
bulk setting imposes on quantum matter. Among other things it allows to focus
on thermal aspects of localization, in particular on those which underlie the
behavior of quantum matter enclosed in black holes. Since QFT is not a theory
of metaphoric miracles, there is a prize to pay in that other aspects, in
particular particles and scattering theory, become blurred in the holographic
projection\footnote{Only in case of wedge-localized algebras in
two-dimensional factorizing models one knows sufficient conditions which allow
to uniquely invert the holographic map.}. Although holography has a priori
nothing to do with gravity, it presents the only known case in which
localization behind causal- and event- horizons may be a physical reality
rather than a theoretical laboratory (Gedankenexperiment) to uncover
interesting structural properties of QFT beyond Lagrangian quantization.

The setting of \textit{wedge-localized algebras} offers the simplest
illustration of the power of holography on null-surfaces. Since the linear
extension of its upper (or lower) horizon is the lightfront, it is not
surprising that the holography in this case may be considered as an autonomous
formulation of the good old metaphoric "lightcone quantization". As its name
indicates lightfront quantization (its more appropriate name) was invented as
a different quantization and hence no attention was paid how it really links
up with the original bulk QFT; not even in the case of free fields one finds
clear statements about its relation to bulk matter and in the context of
interactions its metaphoric aspect was rendering any credible use within the
conceptual setting of QFT impossible. Lighfront holography not only explains
why lightfront quantization failed to be applicable to interacting QFT, but
also saves some of its physical motivations (simplifications in the
description of certain properties). The reason for the failure of the l.c.
quantization approach was that, although in case of free fields there is still
a linear relation to a corresponding generating field of the lightfront
algebra, the presence of interactions destroys any such linear relation. The
only way of relating the bulk matter with its holographic projection consists
in using the field-coordinatization-independent algebraic methods provided by
AQFT. With other words \textit{holography cannot be done in the usual
Lagrangian quantization setting} and even Wightman's more general formulation
would be insufficient.

Although the rigorous mathematical description of lightfront holography (which
is based on modular properties of operator algebras) is quite subtle, its
intuitive physical content can be explained in terms of known classical causal
propagation features for characteristic initial data. The starting observation
is that with the exception of 2-dim. conformal relativistic field theories the
characteristic data can only be specified on one lightfront (say on
$x=+t,x_{\perp}$ arbitrary). Specifying data on half the 3-dim. lightfront
(null) plane determines the data within a wedge ($x>t,x_{\perp}$ arbitrary)
whose upper causal horizon is the half-plane. By sliding the wedge into itself
along the unique lightlike direction contained in its (say upper) horizon, one
obtains a family of wedges whose upper causal horizons are sub-half-planes.
But any (semi) compact region on the lightfront which is not semifinfinite in
the x-directions and also two-sided transverse infinite does not cast any
causal shadow at all i.e. such chracteristic data do not have any associated
ambient causal shadow region. The lightfront contains one unique lightray
direction. The so called Wigner "little group" which leaves this lightray
invariant is isomorphic to the 3-parametric subgroup of the Lorentz group, the
Euclidean group $E(2)$; its 2-parametric "translation" subgroup tilts the edge
of the wedge within the lightfront\footnote{The holographic projection of
these\ Wigner "translations" act like $\left(  x_{\perp}\right)  _{i}-x_{0}$
Galilei velocity transformation in which space and time are interchanged.}
(i.e. it changes the wedges, leaving their upper horizon in the same
lightfront). Altogether the transformations which leave the lightfront
invariant form a 7-parametric subgroup of the Poincar\'{e} group. The
classical globally hyperbolic bulk theory has a symplectic inner product and
the relative symplectic complement of the lightlike shifted classical wedge
algebra inside the original wedge-localized classical subalgebra leads to a
classical subalgebra which is localized in a "slab" of two-sided infinite
transverse extension and compact lightlike extension. Finally the application
of the Wigner translations "tilts" these slabs so that their intersections
define a compact net structure on the lightfront. Modular operator theory
elevates this construction into the algebraic formulation of interacting QFT.

The starting algebra is the wedge-localized algebra $\mathcal{A}(W)\subset
B(H)$ as a subalgebra of all operators in the Hilbert space\footnote{Since the
global lightfront algebra is equal to the global bulk algebra of full
Minkowski spacetime, it cannot be used directly since it violates the \ "no
vacuum annihilator "requirement of the Reeh-Schlieder theorem (the
prerequisite for the application of modular theory).}. The slab-localized
algebras are obtained as relative commutants of the lightlike translated wedge
algebra $\mathcal{A}(W_{a})$ within the original algebra $\mathcal{A}%
(W_{a})^{\prime}\cap\mathcal{A}(W)=\mathcal{A}(slab);$ this resolves the
localization structure in the lighlike direction$.$ Finally the intersection
of these slab algebras with their tilted image under translations in the
little group (by application of the Wigner little group translations) and
their intersections define a local net structure. In this way both the
longitudinal (lightray) as well as transverse localization structure is
resolved. As was already apparent in the classical case of characteristic data
the so-obtained local structure of the holograpgic lightfront projection is
very different from that of the bulk; both structures are local in their own
right, but nonlocal relative to each other\footnote{As mentioned before, the
AdS--CFT correspondence is a less radical type of holography of a AdS bulk
onto a CFT (timelike ) brane for which the relative locality and the maximal
spacetime symmetry is maintained even for the AdS bulk causal shadow
originating from a compact CFT region.}. \ 

Holography can be viewed as an extension of the algebraic isomorphism which in
the new local covariance setting of QFT is the algebraic functorial image of
the diffeomorphism covariance. This means that if one organizes the same
abstract algebraic substrate on two different spacetimes, such that two
globally hyperbolic submanifolds in the different spacetimes happen to be
isometric, then the associated subalgebras are isomorphic i.e. physically
indistinguishable (the quantum theory which would make them also
mathematically identical is the still elusive QG). Null-surfaces have some
common universal features and so do the extended chiral theories which are the
algebraic targets of the holographic projections. In this case an
(holographic) inversion is generally non-unique\footnote{An exception are the
previously mentioned two-dim. factorizing models where one knows the ambient
(bulk) Poincar\'{e} transformation properties of generators ($\simeq$
generators of the Zamolodchikov-Faddeev algebras) of the lightray algebra.}
and one needs additional information to achieve uniqueness.

The (here surpressed) mathematical concept from modular operator theory are
\textit{modular inclusions} for resolving the longitudinal localization and
\textit{modular intersections} for achieving transverse localization.

The resulting local net structure on the lightfront is very interesting,
because the inexorable \textit{vacuum fluctuations} of QFT have all been
\textit{compressed into the longitudinal lightray} direction whereas there are
none in the transverse direction. It is precisely this absence of transverse
vacuum fluctuations which leads to an (transverse) area proportionality of
those quantities (entropy,energy...) which in heat bath thermal setting used
to be be intensive quantities.

It is well known that the restriction of the vacuum state to localized regions
(as e.g. the wedge) is a KMS thermal state at a fixed temperature whose value
depends on how one normalizes the \textit{modular Hamiltonian}; in the case of
the wedge this is the boost generator (of the wedge-preserving boost). KMS
thermal states are well-known to fulfill the second thermodynamic law in its
most general formulation (no perpetuum mobile of second kind) even before a
quantitative notion of entropy has been introduced.

The holographic projection greatly facilitates the computation of the details
about the localization entropy which is associated with the wedge-restricted
vacuum i.e. the proportionality coefficient in the area behavior. The
localization-caused \textit{vacuum entropy of a sharply localized algebra is
formally infinite for the same reasons that the entropy of the
(translation-invariant) global bulk algebra in a heat-bath thermal KMS state
diverges.} Both algebras are what we previously called monades (hyperfinite
type III$_{1}$ von Neumann factors), and in the standard treatment of
heat-bath thermal behavior it is well known that one obtains the volume
proportional entropy formula by approximating the translation-invariant monade
with a sequence of finite box Gibbs states (the famous thermodynamic limit
which defines thermodynamic equilibrium). The same idea works for the
localization entropy, except that in this case the sequence consists of Gibbs
states on fuzzy-localized approximands (type I factors) which converge to the
KMS state on the sharp localized algebra from the inside of the localization
region. Since the argument involves conformal transformations, the metric
difference between long and short distances become irrelevant.

Using the holographic representation of the sharp localized algebra, and
introducing a lighlike \textit{interval of size }$\varepsilon$\textit{ into
which the vacuum polarization cloud of the approximands can spread}%
\footnote{This is not a standards short distance singularity in the sense of
correlations between field-coordinatizations!}, one obtains the following
limiting logarithmic divergence behavior for the localization entropy%
\[
S_{loc}=\lim_{\varepsilon\rightarrow0}Ac\left\vert \ln\varepsilon\right\vert
\]
where $A$ is the area\footnote{The product $A\left\vert \ln\varepsilon
\right\vert $ is precisely the volume factor of a \textit{heat bath system on
the lightfront} whose Hamiltonian is the generator of the lightray
translation. The isomorphism which maps this heat bath system (at $\beta=2\pi
$) to the vacuum-restricted localized one carries the thermodynamic limit
\ into the inner approximand limit and maps the the longitudinal length factor
$L$ in $V=AL$ (conformally) into the $\left\vert ln\varepsilon\right\vert .$}
(of the edge of the wedge) and $c$ is a constant which measures the degrees of
freedom of the holographically projected matter; in typical cases $c$ is equal
to the Virasoro algebra constant. It turns out that the area behavior as well
as the $\left\vert \ln\varepsilon\right\vert $ increase is a totally universal
aspect (as was mentioned in the previous footnote) of localized quantum
matter; its universality is linked to that of an auxiliary global heat bath
thermal system on the lightfront (this isomorphism plays an important role in
the derivation of the formula \cite{AreaI}\cite{AreaII}). Changing the end
points of the smaller interval on the lightray the entropy changes the
ln-factor multiplicatively by the logarithm of the harmonic ratio of the 4 points.

On the other hand, as first observed by Bekenstein, the area behavior which
one encounters in certain classical field theories of the Einstein-Hilbert
kind is more special and has nothing to do with vacuum fluctuations. As
already stated before, a computation of entropy based on quantum mechanical
level counting is (unlike the present localization-entropy) inconsistent with
the local covariance principle.

This conceptual insight is of relevance for the ongoing discussion about
entropy of black holes. In the case of Schwartzschild black holes there are
two ways in which the Hawking effect has been presented.

One way is to take as the relevant state the restriction of a certain static
ground state on the extended matter algebra (which lives on the extended
Schwartzschild spacetime). Upon restriction to the spacetime outside the black
hole appears as a thermal KMS state. In this description the localization
entropy is the unique entropy which the rules of quantum statistical mechanics
relates with the thermal Hawking situation since the Hamiltonian associated
with the Killing symmetry (which has continuous spectrum and therefore admits
no Gibbs state) has to be approximated by a sequence of discrete spectrum
Gibbs states Hamiltonians; this is precisely what the described sequence
achieves. This \textit{QFT in CST setting of the Hawking effect is
conceptually quite tight} and does not offer direct support for speculative
uses of black hole physics towards the still elusive quantum gravity.
\textit{The Hawking radiation is matter radiation without any involvement of
gravitons, it is fully understood in the setting of QFT in CST and the entropy
is the localization entropy associated with this situation.} To define entropy
in any other way which is not related to the approximation of the modular
Hamiltonian does not make much sense since it is only another aspect of the
thermal manifestations of the Hawking effect. Whether it can be used to obtain
hints about QG remains to be seen.

The second way of describing the Hawking effect is the more physical one.
Instead of an equilibrium state one works with a non-equilibrium stationary
state which models a collapsing star. Fortunately the Hawking radiation at
future lightlike infinity does not depend on details, one only needs the to
know the short-distance behavior of the matter two-point function (for free
fields) at the point where the star radius passes through the Schwartzschild
radius. This was Hawking's intuitive idea which later was converted into piece
of beautiful mathematical-conceptual physics in a paper by Fredenhagen and
Haag \cite{Fr-Ha}. Although there has been significant progress in operator
algebra theory on defining an \textit{entropy flux in stationary
non-equilibrium states} which replaces the equilibrium entropy, the
application to the entropy issue of black holes still remains as an
interesting open problem.

The reason for mentioning these ongoing investigation is obvious. The area of
black hole entropy has been under intense investigation for its possible links
to QG. Although I sympathize with taking big jumps into the conceptual blue
yonder in certain situation, I strongly believe that this should not be done
without securing a firm basis from where one could start such a leap (and to
where one could return, if necessary). I do not think that ST is able to
provide such a basis since it itself was founded on very muddy grounds
\cite{Crisis}\cite{decon}. The energy and entropy concept used in ST is a
global one in which quantum mechanical levels are filled and e.g. the entropy
results from counting. As mentioned before this violates the local covariance
principle of QFT in CST which underlies the more dynamical concept of
\textit{localization entropy}. On the other hand the LQG approach is far away
from localizable particle physics and at least in its present form does not
seem to animate particle physicists. As a particle physicist I am very
surprised that the impressive progress about a new QFT framework which not
only incorporates the new local covariance principle \cite{Br-Fr-Ve} but also
promises to shad new light on diffeomorphism invariance \cite{Br-Fr} received
so little attention.

In my mainly verbal presentation I glossed over two intermediate concepts
which are indispensable for the derivation of the above formula. One is what
Nicolov and Todorov \cite{Ni-To} in a recent paper very appropriately called
the "covariant box" Gibbs state.

The other is a very deep analog of the so-called Nelson-Symanzik duality
within the Euclidean Osterwalder-Schrader setting\footnote{Sorry for having to
bring up important milestones of QFT development from the past, which most of
the younger QFTists probably never came across.} \cite{S2}. Whereas in the N-S
duality (for 2-dim. massive QFT in a periodic box place into a KMS thermal
state) the spatial periodicity is simply interchanged with the temperature
"periodicity", the correlation functions of a rotational KMS state on a chiral
algebra turn out to be \textit{selfdual} under interchange of the temperature
with its dual (and a temperature-dependent re-scaling of the angular
coordinates). The case of the one-point function (the thermal expectation of
the identity operator) coalesces with Verlinde's duality for the partition
function for which geometric arguments for its validity were proposed. It
turns out that by viewing the partition function as an object of a complete
thermal QFT one can use the powerful modular operator algebra theory to prove
the temperature duality relation \cite{S1}\cite{S2}. Since this duality
relation is best known under its mathematical name SL(2,$\mathbb{Z}$)
\ "modular identity"\footnote{This terminology refers to the properties of
modular forms as studied in the first half of the 20th century by
mathematicians as Hecke and others.}, what looked like a coincidence of
terminology has now a profound intrinsic connection.

\section{\textit{Holographic symmetry and its relation to the classical BMS
group}}

Another quite amazing consequence of null-surface holography is the emergence
of a gigantic \textit{holographic symmetry group \cite{AreaII}}. If we denote
the generating pointlike fields of the lightfront algebra by $A_{i}(x_{\perp
},x),$ with $x_{\perp}$ being the transverse and $x$ the longitudinal
(lightray) coordinates, their commutation read for such generators
read\footnote{Algebraic CFT, in particular algebraic chiral conformal theory
always posseses pointlike covariant field generators \cite{Joerss}. There can
be no reasonable doubt that this continues to hold for the transverse extended
chiral theories which result from holography.}%
\[
\left[  A_{i}(x_{\perp},x),A_{j}(x_{\perp}^{\prime},x\prime)\right]
=\delta(x_{\perp}-x_{\perp}^{\prime})\sum_{l(k)\geq1}c_{k}\delta
^{l(k)}(x-x\prime)A_{k}(x_{\perp},x)
\]
The quantum mechanical transverse delta function comprises the absence of
transverse vacuum fluctuations and appears in all composite field commutation
relation and even in semiglobal objects (i.e. global in the longitudinal
sense); its ubiquitous presence is the reason for the area behavior (by
integration over $x_{\perp}$). The longitudinal scale dimension of the fields
which contribute on the right hand side determines the degree $l(k)$ of the
derivative of the longitudinal delta function according the well-known rules
of dimensional matching. The automorphism group of this transverse extended
chiral algebra is very big since it includes in addition to transverse
Euclidean transformations the infinite group of $x_{\perp}$-dependent chiral
diffeomorphisms\footnote{The (extended) chiral theories which appear in
null-surface holography, unlike chiral components of two-dimensional conformal
models, do not come with an energy-momentum tensor and hence their
diffeomorphism invariance (beyond Moebius-invariance) need a seperate
discussion.} of the circle (in particular those which fix the point infinity).
In case the null-surface is the upper horizon of a Minkowski spacetime double
cone (i.e. the mantle of a frustum), its linear extension is the mantle of the
(backward) lightcone and the transverse delta function in the above
commutation relation is replaced by a delta function on the unit two-sphere
(with the rotation group replacing the 3-parametric Euclidean group).

Parametrizing the mantle of the frustum in terms of spherical and lightray
coordinates which run from $-\infty$ (the apex) to $+\infty$ we see that the
holographic diffeomorphism group contains (unphysical) copies of the Lorentz
group and the so-called \textit{supertranslations.} In the Penrose limit of an
infinitely large double cone these become identical to the generators of the
classical Bondi-Metzner-Sachs group which is a semi-direct product (or cross
product) of the Lorentz group with the supertranslations. We remind the reader
that the BMS group is defined in a geometric way without reference to QFT.
Instead of looking for Killing isometries, one studies the much more general
concept of transformations which fulfill the Killing equation in an
appropriate asymptotic sense. The result is that semidirect product of angular
dependent lightlike translations ("supertranslations") with the Lorentz group.
The existence of a relation to the holographic group is not surprising in view
of the similarity of Penrose's picture for (future) lightlike infinity.

The surprising aspect is the fact that the conceptual relation with respect to
properties of the bulk matter is much deeper in the quantum case. Since
holography is a change of spacetime encoding in the same Hilbert space, the
action of the holographic group is well-defined on the full algebra inasmuch
as the global symmetry of the full algebra (Poincar\'{e}, conformal) has a
well-defined action in terms of a geometric change of the null-surface. The
only subgroup of the holomorphic group which acts as a diffeomorphism on the
full algebra is the Poincar\'{e} group, all other holomorphic symmetries act
in a \ "fuzzy" way which can only be described in terms of algebraic concepts
(in the case of pointlike field coordinates in terms of a transformation on
testfunction spaces). Naturally those quantum symmetries have no associated
Noether theorem.

\section{Conclusions}

In these notes we set out to illustrate the power of the intrinsic algebraic
formulation of QFT in which particle physics is not described in terms of
individual field-coordinatizations but rather in terms of relations between
spacetime-indexed operator algebras. The case of thermal manifestations of
quantum matter behind causal/event horizons is most appropriate for this
purpose because this phenomenon is outside the range of Lagrangian
quantization. But as was indicated in the introduction, AQFT also offers a new
perspective on gauge theories and the Higgs issue. In view of the expected
observational progress expected from the LHC collider it should be very
interesting to elaborate these ideas.


\begin{thebibliography}{99}                                                                                               %


\bibitem {Khar}P. Jordan, \textit{The Present State of Quantum
Electrodynamics}, in \textit{Talks and Discussions of the Theoretical-Physical
Conference in Kharkov} (May 19.-25., 1929) Physik.Zeitschr.XXX, (1929) 700

\bibitem {Haag}R. Haag, \textit{Local Quantum Physics}, Springer 1996

\bibitem {Le}G. Lechner, \textit{An Existence Proof for Interacting Quantum
Field Theories with a Factorizing S-Matrix}, math-ph/0601022

\bibitem {Ba-Fo-Ka}H. Babujian, A. Foerster, M. Karowski, \textit{Exact form
factors in integrable quantum field theories: the scaling Z(N)-Ising model},
Nucl.Phys. B736 (2006) 169, hep-th/0510062

\bibitem {Crisis}B. Schroer, \textit{String theory and the crisis in particle
physics (a Samizdat on particle theory)}{\small , }to be published in
I.J.M.P.D., physics/0603112

\bibitem {AreaI}B. Schroer, \textit{Area density of localization-entropy I:
the case of wedge-localization}, Class.Quant.Grav. 23 (2006) 5227

\bibitem {AreaII}B. Schroer, Area density of localization-entropy II: double
cone-localization, holographic symmetry and the Bondi-Metzner-Sachs group

\bibitem {Su}F. J. Summers, \textit{Tomita-Takesaki Modular Theory}, math-ph/0511034

\bibitem {Ka-Wi}R. Kaehler and H.-W.. Wiesbrock, JMP \textbf{42}, (2000) 74

\bibitem {sub}V. F. R. Jones, \textit{A new knot polynomial and von Neumann
algebras}, Notices of the AMS, March 1986, p.\ 219

\bibitem {Yng}J. Yngvason, Rept.Math.Phys. 55 (2005) 135, math-ph/0411058

\bibitem {F-R-S}K. Fredenhagen, K.-H. Rehren and B. Schroer, Rev. math. Phys.
\textbf{SI}1 (special issue) (1992) 113

\bibitem {PFG}B. Schroer, \textit{Modular Wedge Localization and the d=1+1
Formfactor Program}, Annals Phys. 275 (1999) 190

\bibitem {Wi-St}R. F. Streater and A. S. Wightman, \textit{PCT, Spin and
Statistics and All That}, Benjamin, New York 1964

\bibitem {B-B-S}H. J. Borchers, D. Buchholz and B. Schroer, Commun. Math.
Phys. \textbf{219}, (2001) 125, hep-th/0003243

\bibitem {Ka}M. Karowski and P. Weisz, Phys. Rev. B \textbf{139}, (1978) 445

\bibitem {Bu-Le}D. Buchholz and G. Lechner, \textit{Modular Nuclearity and
Localization}, math-ph/0402072

\bibitem {Bu}D. Buchholz, \textit{On the manifestations of particles}, hep-th/9511023

\bibitem {Ho}G. 't Hooft, in Salam-Festschrift, A. Ali et al. eds., World
Scientific 1993, 284

\bibitem {Mal1}J. M. Maldacena, \textit{Large N-limit of superconformal field
theories and supergravity}, Adv. Theor. Math. Phys. 2, (1998) 231

\bibitem {Dies}http://golem.ph.utexas.edu/\symbol{126}distler/blog/archives/000987.html

\bibitem {Mal2}J. M. Maldacena, \textit{The Illusion of Gravity}, \ Scientific
American, (November 2005) 56

\bibitem {S1}B. Schroer, \textit{Two-dimensional models as testing ground for
principles and concepts of local quantum physics}, Annals Phys. 321 (2006)
435, hep-th/0504206

\bibitem {S2}B. Schroer, \textit{Positivity and Integrability (Mathematical
Physics at the FU-Berlin)}, hep-th/0603118

\bibitem {Fr-Ha}K. Fredenhagen and R. Haag, Commun. Math. Phys. \textbf{127},
(1990) 273

\bibitem {decon}B. Schroer, \textit{String theory deconstructed} {\small (a
detailed critique of the content of ST from an advanced QFT viewpoint), }in preparation

\bibitem {Br-Fr-Ve}R. Brunetti, K. Fredenhagen and R. Verch, \textit{The
generally covariant locality principle -- A new paradigm for local quantum
physics}, Commun.Math.Phys. 237 (2003) 31, math-ph/0112041

\bibitem {Br-Fr}R. Brunetti and K. Fredenhagen, \textit{Towards a Background
Independent Formulation of Perturbative Quantum Gravity}, gr-qc/0603079

\bibitem {Ni-To}Nikolay M. Nikolov, Ivan T. Todorov, \textit{Lectures on
Elliptic Functions and Modular Forms in Conformal Field Theory}, math-ph/0412039

\bibitem {Joerss}M. J\"{o}rss, \textit{From Conformal Haag-Kastler Nets to
Wightman Functions}, Lett. Math. Phys. \textbf{38}, (1996) 257, hep-th/9609020
\end{thebibliography}
\end{document}